\documentstyle[11pt,ska,twoside,psfig]{article}
\begin{document}
\title{Late stages of stellar evolution with the Square Kilometer Array}
 \author{Albert A. Zijlstra}
\affil{UMIST, Department of Physics, P.O. Box 88, Manchester M60 1QD, UK}


\begin{abstract}
Stars at the end of their lives return much of their mass back to the
ISM.  This process is the main source of dust in the ISM, and
potentially a source of large molecules. I discuss several areas where
SKA will have major impact on our understanding of the mass loss and
the circumstellar processes. The resulting requirements for SKA,
including frequency coverage and location, are discussed.
\end{abstract}

\section{Introduction}

Low and intermediate-mass stars, near the end of their lives,
experience a phase of catastrophic mass loss. During this so-called
superwind, occuring on the Asympotic Giant Branch (AGB), between 40\%\
and 80\%\ of the initial mass of the star is ejected at mass-loss
rates of typically $10^{-6 \rightarrow -4}\,\rm M_\odot\,yr^{-1}$.
The total amount lost determines the ISM recycling rate: the fraction of
the gas incorporated in star formation which after a delay of
$10^{8--10}\,\rm yr$ is returned to--and rejuvenates--the ISM. In the
Galaxy, this rate is of the order of $0.5\,\rm M_\odot\,yr^{-1}$.
The ejecta are enriched in carbon and nitrogen; at low metallicity,
primary oxygen is also produced. The AGB superwind is the site of
efficient dust production, and much of this dust survives the later
evolution to become part of the ISM. In addition to dust, complex
organic molecules can also be formed around AGB stars. If these can
survive normal ISM conditions (a topic of current research), AGB stars
form a source of pre-biotic molecules into regions of later star
formation. Finally, the mass-loss efficiency determines the final
stellar mass and is the cause that all stars with $M_{\rm initial} <
8\,\rm M_\odot$ form white dwarfs rather than supernovae.

The superwind strongly affects the evolution of
galaxies. It determines the lower
limit for supernova masses.  If, as expected, the superwind is less
efficient at low metallicity, this would increase the masses of
stellar remnants, and increase the supernova rate in the early Universe.
It also predicts two sharp changes in the
composition of newly formed dust. The first occurs after $10^8$ yr
when the first AGB stars appear. The second follows when the initial main
sequence turn-off mass drops below the value (1.5--4\,$\rm M_\odot$)
where AGB stars no longer become carbon stars, and form silicate rather than
carbonaceous dust.

Unsolved issues regarding the AGB superwind include the
structure of the wind (morphology, clumps, embedded magnetic fields),
and the chemistry and dust formation. The SKA is well suited to 
studying these: it can resolve the extended envelopes 
down to the radio photosphere, and study a wide range of emission 
processes.

\section{Radiation processes}

During the superwind, the star is a red giant with a radius of around
1 AU and a temperature of around 3000\,K. The extended atmosphere is
dominated by radially travelling shock waves, and shows extensive
chemistry and dust formation.  Optically, the
star and its atmosphere tend to be obscured ($A_V \sim 10--30\,\rm
mag$). The radio photosphere is located within the extended atmosphere.
After the mass loss ceases, the star heats up very fast and ionizes
the ejecta, now visible as a planetary nebula (PN). Depending on the
precise evolutionary phase, the SKA will detect radio emission from the
following processes:

\begin{enumerate}

\item Thermal bremsstrahlung, from the ionized gas of planetary
nebulae.  This will have a flat $\nu^{-0.1}$ spectrum, except for the
youngest objects which are optically thick at low frequencies. For a
sensitivity level of 0.1$\mu$Jy, the bremsstrahlung can be detected to 
5\,Mpc for the brightest objects. A typical young PN (e.g NGC 7027)
can be resolved to about 1\,Mpc, assuming observations at 8\,GHz
with 10$\mu$arcsec resolution (1000\,km baseline).

\item Thermal dust emission. Within the radio frequencies, this is
normally much weaker than the bremsstrahlung. To detect this in PNe, a
high dynamic range of 1000:1 will be needed, and frequency coverage
should extend to $\nu > 10$GHz. However, in AGB stars it may
contribute significantly.

\item Photospheric continuum emission, from the AGB stars. For a
sensitivity of 0.1$\mu$Jy at 20\,GHz, this can be detected to 3--4\,
kpc, covering around $10^4$ potential targets. At a resolution of 5
$\mu$arcseconds, the radio photospheres (3--5\,AU in diameter) can be
resolved out to 1--2 kpc.  Because of the black-body nature, the
preferred frequency range is 10--20\,GHz for the optimum combination
of resolving power and sensitivity. Nothing is known about radio flares from
AGB stars, but these are likely to exist.

\item Molecular transitions, both masing and non-masing. The
well-known OH 18-cm lines from AGB stars can be detected throughout
the Local Group. Here we assume a typical OH component of 10\,Jy over
$0.5\,\rm km\,s^{-1}$, for a source 1\,kpc away.  At this resolution
the line sensitivity is about $10^3$ less than the continuum
sensitivity. The OH line is detectable out to 2.5\,Mpc and OH/IR AGB
stars can be studied throughout the Local Group.  SKA will also be
able to study all four lines simultaneous.  The water masers at 22-GHz
are equally important. Other important molecular lines are mainly at
frequencies $\nu>10$GHz and are detectable for Galactic stars.

\item Radio recombination lines. These are potentially the best lines
to get detailed velocity fields, and they also give good abundance
determinations as the recombination lines are not affected by
extinction, and lack the extreme temperature dependence of (optical) forbidden
lines. To observe a range of lines tracing a range of conditions, the
frequency range should go up to $\sim 20\,$GHz.  Assuming a 5\%\
line-to-continuum ratio, a bright PN can have a line flux of 0.1\,mJy
at 200\,kpc, measurable at a velocity resolution of $10\rm
\,km\,s^{-1}$. Recombinations lines from heavier elements can also be
detected. Faint lines for deuterium and $^3$He may be especially
interesting.

\item HI. A young planetary nebula will still be surrounded by a
photon-dissociation region, including a larger atomic shell. The young
PN NGC\,7027 has a circumstellar HI optical depth of around 0.05. The
absorption line is detectable to same distance as the recombination
lines. They can be used to measure magnetic fields, given the S/N
requirements mainly within the Galaxy.

\end{enumerate}
 
The
sensitivity of line emission does not greatly benefit from the new
wide-band correlators, such as used for e-VLA (although the continuum
will be much better determined, and for weak lines this does increase
the S/N). The large collecting area of SKA is a requirement for 
detecting faint lines.

The above assumes nominal continuum sensitivity. However, the
continuum sensitivity within the galactic plane may be limited by
confusion. The stellar density for a line of sight through the plane
may be of the order of 10 stars per square arcsecond. Radio emission
from stars is generally weak but can rise dramatically during strong
flares, especially among the numerous M-dwarfs.  Thus the background
continuum emission is likely to contain a rapid time-variable
component.

The numbers given take into account that the sensitivity will depend
on frequency and resolution. At high frequencies, the effective
collecting area may be reduced. Also, much of the collecting area will
be concentrated at shorter baselines. The highest spatial resolution
will involve a subset of the array only.

\section{Winds, shells and chemistry}

\subsection{Clumps in planetary nebulae}

Planetary nebulae show many varied structures, including rings, jets,
cometary structures and clumps. How much of these are directly related
to structures in the original AGB wind is not known. Especially the
SiO masers show that the AGB wind is intrinsically clumpy. If these
clumps survive to become the PNe clumps, they could trace regions with
distinct evolution. 

Recent analyses of the heavy element, forbidden and recombination line
spectra of planetary nebulae (PNe) suggest that there are at least two
distinct emission regions in many nebulae, one of `normal' temperature
($T_e \sim 10^4\,$K) and `normal' abundances where the strong
collisionally excited forbidden lines originate; and another which is
inferred to have a very low temperature ($10^3$\,K) and very high
heavy elemental abundances, which emits most of the observed flux from
the heavy element optical recombination lines but emits essentially no
forbidden line emission (see the recent review by Liu, 2002).  It is
tempting, but still unproven, to associate the second component with
some of the clumps.

The thermal continuum emission is ideally suited to locate the
proposed super-metal-rich gas.  The inverse dependence on temperature
of free-free continuum emission ($1/\sqrt{T_e}$ in the optically thin
case) makes these clumps three times brighter than clumps with normal
temperature.  We assume a size for a clump of $10^{15}\,$cm and
density of $10^4$: the mass of the clump is of the order of $5 \times
10^{-8}\,\rm M_\odot$ and the size 0.1 arcsec at 1\,kpc. The radio
flux density from the clump is about 1\,$\mu$Jy (or a little fainter
if hydrogen is underabundant.) This is within SKA territory: it will
be able to detect individual clumps with such parameters.  HST images
show that dust in the nebula is causing extinction at very small
angular scales. In case of super-metal-rich clumps, an association
with dust enhancements is probable.  SKA images, not affected by
extinction, when compared to AO images, will show the internal dust
distribution:

Recombination lines are highly important: they show a stronger
dependence of $1/T_e$ and can obtain the temperature of the gas.
Lines from individual clumps will not be detectable, but summed over
the clumps, the integrated line profile would be observable. The SKA
bandwidth will cover several recombination lines simultaneously. The
line strength as function of upper level $n$ gives an independent
measurement of the density of the gas. In addition, velocity fields
can be measured. For the first time, a full understanding of the
nebulae structure will be obtained.  Extreme hydrogen-poor gas (as is
observed in a few PNe, e.g. A58) would give strong recombination lines
from element such as carbon, giving a new tracer for these unusual
environments.

\subsection{The proto-planetary nebulae}

One of the unsolved problems in evolved stars is the origin of the
asymmetrical morphologies. Stellar winds are thought to be largely
spherical. However, detached circumstellar shells tend to be
elliptical or bipolar--e.g. $\eta$ Car. There are presently about 100
objects known in the phase between the AGB and the PNe, where the shell
is already detached but the star not yet hot enough to initiate ionization.
These proto-planetary nebulae (PPNe) are almost always highly aspherical,
with evidence for dense disks and fast polar outflows. The origin of these
structures is not understood, as their immediate progenitors on the AGB 
show little or no evidence for disks. Suggestions range from infall of
planetary companions (with their angular momentum deposited in the wind),
to magnetic fields originating in the turbulent convection of the AGB
stars. 

The youngest outflows can be observed in several maser transitions of
OH and H$_2$O. (The 43GHz SiO masers will be outside the frequency
coverage of SKA.) Zijlstra et al. (2001) have shown that the OH
velocity fields of some nearby objects show the characteristic trace
of an interaction between winds of different speeds. This results in a
swept-up intershell moving at a direction-dependent velocity, which is
constant in time.  A sharp increase in the wind speed, possibly
already on the AGB, may therefore trigger the change in
morphology. SKA will be able to resolve the problem in various ways:

\begin{itemize}
\item Spatial resolution. The VLA, used to obtain sufficient S/N on
long baselines, limits the spatial resolution to about 1
arcsecond. SKA will improve the spatial resolution by a factor of 30
at high S/N.  This region, within 10\,AU of the star,
will show the onset of the acceleration on the AGB.

\item Expansion. Brighter maser spots will have their position measured
to 1 milli-arcsecond.  Assuming an expansion at 30\,km\,s$^{-1}$,
the spatial evolution can be detected within 5 years out to 30\,kpc. 
The evolution of the structures can be directly measured. 
The same method will determine the distance to the Galactic Centre,
the Sgr dwarf and the two Magellanic Clouds. However, we expect that this
will be measured before SKA by other methods.

\item Magnetic fields. Especially OH has a large Zeeman
splitting. Identifying Zeeman components gives a measurement for the
magnetic field, if one can show that the two components are
co-spatial. The 6-cm OH lines are in principle better than the 18-cm
ones, giving a larger splitting and better resolution. At present,
these have only been detected in a very few objects, but SKA will
detect them widely. Provided circular polarisation can be accurately
calibrated, SKA will clean up the problem of magnetic field
structures in these outflows.

\end{itemize}

\subsection{Chemistry}

AGB stars become enriched in carbon during the third dredge-up. For
intermediate-mass stars, the carbon will eventually outnumber oxygen.
In these stars, the photosphere contains C$_2$ and C$_2$H$_2$, which
are carried along with the mass loss.  Two separate chemical sequences
occur: carbon chains build up and eventually become
aromatic, while other molecules cluster and form the first
condensates. The present evidence is that carbon dust forms from
TiC$_2$, which has a high condensation temperature. The aromatic
molecules build up to PAHs. Earlier suggestions that the PAHs form
dust are not born out by models, however they may become incorporated
into dust grains.

The evidence is growing that both the dust and molecules survive the
stellar evolution and merge into the ISM. For the dust, direct evidence 
comes from the inclusions into meteorites. Of the identifiably pre-solar
grains found in the solar system, the majority show an AGB origin.
This suggests that most of the interstellar dust in the ISM formed
in AGB envelopes, and not, as previously believed, in supernovae.

The diffuse ISM also shows a large number of absorption bands (DIBs)
which are unidentified but appear to come from large carbon molecules.
These must be sufficiently large to be stable against
photo-dissociation, and so must be considerably larger than a small
PAH. Their origin is unknown. However, given the origin of the dust in
the ISM, it is plausible that these molecules also originate from AGB
stars. They may either have grown to photo-stable sizes in the gaseous
envelopes, or they may have become incorporated in dust and evaporated
from dust grains at a later phase.

The first step in forming aromatic molecules is benzene
(C$_6$H$_6$). Once this forms, larger rings easily build up.  Wood et
al. (2002) have shown an efficient chemical route to this molecule,
operating in dense, irradiated post-AGB envelopes. This suggests that
PAHs do not form during the AGB, but are a later product. The
important questions are how (and where) the chemistry is initiated,
and where the chemistry ends. Proof of the formation of aziridine
(c-CH$_2$CH$_2$N) would be important. Furan (Dickens et al. 2001) is
also expected to exist, and is the basis of the simple sugars ribose
and deoxyribose, involved in building RNA and DNA.  Glycine, the
simplest amino-acid, has been a target for searches (Charnley et
al. 2001).  If such molecules form and survive into the ISM, they also
become part of any subsequent star formation. The pre-solar nebula
would in this way have become pre-seeded with organic molecules. Amino
acids have been detected in meteorites, suggesting a solar-nebula or
pre-solar origin.

SKA can detect and identify such molecules. The lines are faint and
the large collecting area of SKA would increase the sensitivity by
orders of magnitude. In addition, there are fewer transitions in the
SKA frequency range, and these can be measured in the laboratory.
This allows an identification of the lines. ALMA, in contrast,
observes in a frequency range where there are many overlapping
transitions, and identification is problematic. In spectroscopic
sub-mm surveys, typically only the brightest lines can be identified
and the 'noise' consists of many weak unidentified lines. SKA will have
a major impact in the rapidly developing field of pre-biotic ISM
chemistry.

The best region to get a mix between a sufficient range of molecules
detectable without the problems of confusion, is found between 10 and
30 GHz. Furan, as an example, has lines at 23.4\,GHz and glycine
(e.g. Lovas et al. 1995) at 22.7\,GHz.




\section{SKA requirements}

The sensitivity is taken from the defined SKA parameters. The spatial
distribution of the array determines how this changes
at the highest spatial resolution. If a trade-off exists between
multi-beam and frequency coverage, one can envisage that the highest
frequencies will only be offered in a sub-array. For spectral line
observations, it is desirable to cover a large number of lines
simultaneously, where each part of the spectrum must allow for high
resolution, while at the same time the continuum level can also be
accurately determined.

Good frequency coverage is crucial.  Free--free emission is best
studied at higher frequencies, especially for high
density, optically thick regions.  The OH lines are at 1.6 and
6\,GHz. The water masers are at 22\,GHz. The large carbon molecules
make an upper frequency of order 30\,GHz highly desirable.  A
frequency gap between  ALMA  and SKA performance should be
avoided.

Finally, the  majority of the
Galactic and near-Galactic environments require observations from the
Southern hemisphere. It is also desirable to be able to observe the
same targets as ALMA. Accommodating an array of up to 1000\,km,
able to observe e.g. the Galactic Centre region,
suggests a location in Australia or South Africa would maximise the scientific
return.


\end{document}